\pdfoutput=1

\documentclass[
  aps,
  prl,
  reprint,
  amsmath,
  amssymb,
  superscriptaddress,
  longbibliography,
  floatfix
]{revtex4-2}

\usepackage{graphicx}
\usepackage{bm}
\usepackage{mathtools}
\usepackage{microtype}
\usepackage[colorlinks=true,allcolors=blue]{hyperref}

\newcommand{\dd}{\mathop{}\!\mathrm{d}}
\newcommand{\e}{\mathrm{e}}

\newcommand{\Var}{\operatorname{Var}}
\newcommand{\tlambda}{\widetilde\lambda}
\newcommand{\tOmega}{\widetilde\Omega}

\begin{document}

\title{Scrambling Separates Superradiance from Quantum Chaos}

\author{Long-Jie Li}
\affiliation{Institute of Quantum Physics, Hunan Key Laboratory of Nanophotonics and Devices, Hunan Key Laboratory of Super-Microstructure and Ultrafast Process, School of Physics, Central South University, Changsha 410083, China}

\author{Li-Li Gao}
\affiliation{Institute of Quantum Physics, Hunan Key Laboratory of Nanophotonics and Devices, Hunan Key Laboratory of Super-Microstructure and Ultrafast Process, School of Physics, Central South University, Changsha 410083, China}

\author{Zi-Cheng Liu}
\affiliation{Institute of Quantum Physics, Hunan Key Laboratory of Nanophotonics and Devices, Hunan Key Laboratory of Super-Microstructure and Ultrafast Process, School of Physics, Central South University, Changsha 410083, China}

\author{Zi-Min Li}
\email{zimin.li@csu.edu.cn}
\affiliation{Institute of Quantum Physics, Hunan Key Laboratory of Nanophotonics and Devices, Hunan Key Laboratory of Super-Microstructure and Ultrafast Process, School of Physics, Central South University, Changsha 410083, China}

\author{Murray T. Batchelor}
\affiliation{Mathematical Sciences Institute, Australian National University, ACT 2601, Australia}

\date{October 20, 2026}

\begin{abstract}
Early studies on the Dicke model linked superradiance to quantum chaos, whereas later anisotropic Dicke and Rabi studies reached apparently conflicting conclusions.
We resolve this tension by distinguishing global chaos from local instability.
Along the isotropic Dicke path, a sweep at fixed low excitation energy crosses both the superradiant boundary and a nearby chaotic window, explaining the reported correlation.
However, the anisotropic Dicke model can be chaotic in the normal phase and regular in the superradiant phase, making superradiance neither necessary nor sufficient for global chaos.
A fidelity out-of-time-order correlator (FOTOC), in contrast, probes the local stability of the stationary point selected by the initial wave packet.
The superradiant transition destabilizes the origin, whereas a change in anisotropy can restore its stability and create displaced saddles.
Wave packets centered at these saddles exhibit exponential FOTOC growth even on a classically regular Dicke energy shell with Poisson statistics.
This local sensitivity also enables quantitative spectroscopy: combined with the topology of stationary points and local density of states relations, FOTOC dynamics reconstructs the absolute amplitudes of spectral singularities in both models without fitting the density of states.
\end{abstract}

\maketitle

The Dicke model, introduced for collective emission \cite{Dicke1954}, exhibits a thermodynamic transition between normal and superradiant phases \cite{HeppLieb1973,WangHioe1973,Kirton2019}.
In an influential study, Emary and Brandes found the onset of the superradiant transition to coincide with classical chaos and Wigner--Dyson level statistics \cite{EmaryBrandes2003}.
However, subsequent work on the anisotropic Dicke model found chaos to emerge away from integrable limits, independently of the superradiant boundary \cite{Buijsman2017}.
Yet out-of-time-order correlators (OTOCs) in the Dicke model \cite{LewisSwan2019,AlaviradLavasani2019} and the quantum Rabi model \cite{Rabi1936,LiBatchelorGAA2021} appeared to restore this connection through rapid scrambling near the transition \cite{Kirkova2022}.
Given that the Rabi and Dicke models share the same semiclassical superradiant bifurcation \cite{Hwang2015,Liu2017,Sun2020}, this tension raises a concrete question: does this bifurcation govern global chaos or local instability?

An unstable stationary point can produce early exponential OTOC growth without global chaos \cite{PilatowskyCameo2020,XuScaffidiCao2020,Kidd2021}.
Early exponential growth has also been reported for states centered in both stable and chaotic regions of the anisotropic Rabi model's semiclassical phase space \cite{Wang2025}.
These findings need not contradict the reported connection between scrambling and superradiance, since the diagnostics probe different properties.
Superradiance is a ground-state property, global chaos characterizes a fixed-energy shell, and a localized fidelity OTOC (FOTOC) depends on the initial state and probe.
A systematic comparison is therefore needed to explain when and why these properties are correlated.

Although exponential OTOC growth alone does not establish global chaos, a localized FOTOC may still serve as a quantitative spectroscopic probe.
The local instability rate helps determine the strength of a density-of-states singularity, whose form depends on stationary-point geometry and the number of classical degrees of freedom \cite{Hummel2019,StranskyCejnar2016,Cejnar2021}.
Can local FOTOC dynamics therefore reconstruct the absolute amplitudes of spectral singularities in both models on a common scale?

Here we use the shared stationary-point geometry to compare superradiance, global chaos, and local scrambling in the Rabi and Dicke models.
This comparison shows that superradiance is neither necessary nor sufficient for global chaos on anisotropic Dicke energy shells and explains its apparent connection to chaos along the isotropic path.
By contrast, local FOTOC dynamics tracks the stability of the stationary point selected by the initial state, allowing exponential growth without global chaos.
We further combine FOTOC dynamics with stationary-point geometry and local density-of-states relations to reconstruct the absolute amplitudes of spectral singularities in both models.

\begin{figure*}[t]
\begin{minipage}[t]{\dimexpr(\textwidth-\columnsep)/2\relax}
\vspace{0pt}
\includegraphics[width=\linewidth]{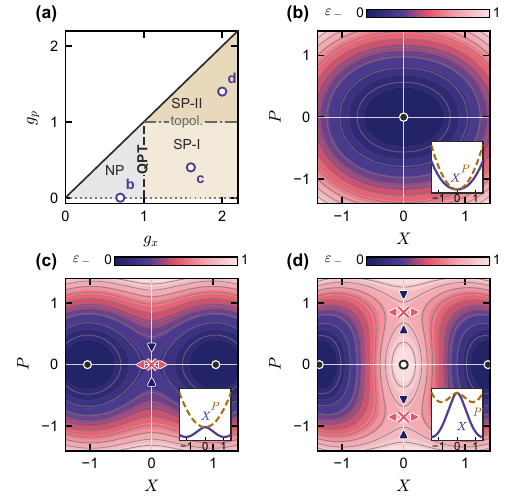}
\caption{\label{fig:landscapes}Stationary-point geometry of the shared ARM--ADM energy surface.
(a) Coupling plane, with points b--d selecting panels (b--d).
(b--d) Separately normalized energy contours at $(g_x,g_p)=(0.7,0)$, $(1.6,0.4)$, and $(2,1.4)$ in NP, SP-I, and SP-II, respectively.
Insets show energy cuts along $X$ and $P$ relative to the origin.
Filled circles, crosses, and open circles mark minima, saddles, and maxima.
Blue and red arrows mark positive and negative curvature directions, respectively.}
\end{minipage}\hfill
\begin{minipage}[t]{\dimexpr(\textwidth-\columnsep)/2\relax}
\vspace{0pt}
\includegraphics[width=\linewidth]{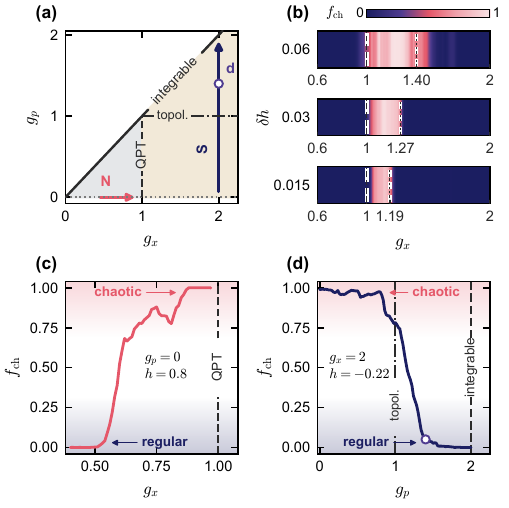}
\caption{\label{fig:global}Ground-state superradiance does not determine global chaos on ADM energy shells.
(a) Coupling paths N (normal phase) and S (superradiant phase).
(b) $f_{\rm ch}$ at $g_p=0$ and fixed excitation energies $\delta h=0.015,0.03,0.06$ above the ground state (bottom to top).
White guides mark $g_x=1$ and $g_x^*$, where $\Delta h_{\rm sep}(g_x^*)=\delta h$.
(c,d) $f_{\rm ch}$ along N and S, respectively.
Open circles in (a,d) mark $(g_x,g_p)=(2,1.4)$.}
\end{minipage}
\end{figure*}

\emph{Models and shared adiabatic surface.---}
The anisotropic Rabi model (ARM) and anisotropic Dicke model (ADM) are described by the common Hamiltonian \cite{Buijsman2017,Liu2017,LiEtAl2024}
\begin{equation}
\begin{aligned}
\hat H_N={}&\omega\hat a^\dagger\hat a+\Delta\hat J_z
+\frac{g_1}{\sqrt N}(\hat a^\dagger\hat J_-+\hat a\hat J_+)\\[-2pt]
&+\frac{g_2}{\sqrt N}(\hat a^\dagger\hat J_++\hat a\hat J_-).
\end{aligned}
\label{eq:HN}
\end{equation}
Here $N$ is the atom number, $\omega$ the field frequency, and $\Delta$ the atomic splitting, with $\Delta=\omega$ for the ADM.
The collective spin operators are $\hat J_\alpha=\frac12\sum_i\hat\sigma_\alpha^{(i)}$, with $\hat J_\pm=\hat J_x\pm i\hat J_y$.
The couplings $g_1$ and $g_2$ describe the rotating and counterrotating interactions, respectively.
Equation~\eqref{eq:HN} gives the ARM for $N=1$ and the ADM in the maximal-spin sector $j=N/2$ for $N>1$.

Using the critical scale $g_c=\sqrt{\Delta\omega}$, we define dimensionless couplings for the position and momentum quadratures, respectively:
\begin{equation}
g_x=\frac{g_1+g_2}{g_c},\qquad
g_p=\frac{g_1-g_2}{g_c}. \label{eq:quadrature-couplings}
\end{equation}
We consider $g_1\geq g_2\geq0$, corresponding to $0\leq g_p\leq g_x$.
Within this region, $g_p=0$ is the isotropic line, whereas $g_p=g_x$ is the $g_2=0$ $U(1)$ integrable limit.
In their respective limits, $N\to\infty$ for the ADM and $\eta=\Delta/\omega\to\infty$ for the ARM, both models exhibit the superradiant quantum phase transition (QPT) at $g_x=1$ \cite{Hwang2015,Liu2017,Peng2019}.

Minimizing over the spin orientation at fixed scaled field quadratures $X,P$ \cite{SupplementalMaterial} yields the same reduced lower-branch energy surface in both models,
\begin{equation}
\varepsilon_-(X,P)=\frac{X^2+P^2}{2}
-\frac12\sqrt{1+2g_x^2X^2+2g_p^2P^2}. \label{eq:surface}
\end{equation}
However, this shared stationary geometry does not imply equivalent dynamics.
The adiabatic ARM has one classical canonical pair, whereas the ADM mean-field phase space retains both field and collective-spin pairs.
The additional pair therefore allows nonintegrable motion on an ADM energy shell.

\emph{Stationary-point geometry.---}
For $0\leq g_p<g_x$, the stationary points of the shared surface distinguish the normal phase (NP) from two topologies within the superradiant phase, SP-I and SP-II [Fig.~\ref{fig:landscapes}(a)].
For $g_x<1$, the origin is the only minimum [Fig.~\ref{fig:landscapes}(b)].
Crossing $g_x=1$ at $g_p<1$ creates two $X$-axis minima and turns the origin into a saddle, defining SP-I [Fig.~\ref{fig:landscapes}(c)].
Within the superradiant phase, a second change occurs at $g_p=1$: the quadratic term in $P$ vanishes and the energy becomes locally quartic in $P$.
Beyond this boundary, the origin becomes a dynamically stable maximum, while two saddles $P_\pm$ appear on the $P$ axis, defining SP-II [Fig.~\ref{fig:landscapes}(d)] \cite{HeSong2026}.
Thus, $g_x=1$ marks the onset of superradiance, whereas $g_p=1$ relocates the local instability within the superradiant phase.

\emph{Superradiance does not determine global chaos.---}
We quantify global chaos on ADM energy shells labeled by $h=E/(2N\Delta)$, with $\epsilon=2h$ for ARM--ADM comparisons.
We characterize each shell by its chaotic fraction $f_{\rm ch}$, the fraction of sampled initial conditions classified as chaotic using finite-time Lyapunov exponents \cite{SupplementalMaterial}.
Here $f_{\rm ch}=0$, $1$, and intermediate values denote regular, fully chaotic, and mixed energy shells, respectively.
$f_{\rm ch}$ characterizes the energy shell as a whole, in contrast to a localized FOTOC.

To understand the reported correlation between superradiance and chaos \cite{EmaryBrandes2003}, we follow the isotropic line $g_p=0$ at fixed excitation energy $\delta h\equiv h-h_0(g_x)$ above the ground-state energy $h_0(g_x)$ [Fig.~\ref{fig:global}(b)].
For $g_x>1$, the energy difference between the origin saddle and the superradiant minima gives the barrier between the two wells,
\begin{equation}
\Delta h_{\rm sep}=\frac{(g_x-g_x^{-1})^2}{8}. \label{eq:barrier}
\end{equation}
At fixed $\delta h$, interwell motion requires $\delta h\geq\Delta h_{\rm sep}$ and becomes inaccessible beyond $g_x^*$, defined by $\Delta h_{\rm sep}(g_x^*)=\delta h$.
However, $f_{\rm ch}$ remains nonzero beyond $g_x^*$, showing that trajectories confined to a single well can still be chaotic.
At these low excitation energies, the chaotic window lies near the superradiant boundary, accounting for the reported correlation along the isotropic path.

To test whether this correlation is general, we now vary the couplings without crossing the superradiant boundary.
Along path N in Fig.~\ref{fig:global}(a) ($g_p=0$, $0.4\leq g_x\leq0.9$), the ground state remains normal, yet $f_{\rm ch}$ rises from nearly zero to unity on the shell $h=0.8$ [Fig.~\ref{fig:global}(c)].
At the two endpoints, the parity-resolved adjacent-gap-ratio distributions likewise change from Poisson to Gaussian orthogonal ensemble (GOE) statistics \cite{Atas2013,SupplementalMaterial}.
Superradiance is therefore not necessary for global chaos.

Conversely, along path S in Fig.~\ref{fig:global}(a) ($g_x=2$, $h=-0.22$), $f_{\rm ch}$ falls from nearly unity to zero as $g_p$ approaches the integrable line, although the ground state remains superradiant [Fig.~\ref{fig:global}(d)].
Together, paths N and S establish that superradiance is neither necessary nor sufficient for global chaos.

\emph{Why Rabi scrambling can track the QPT.---}
In the ARM, a localized wave packet probes the stability of the stationary point at its center.
For a kick generated by a quadrature $G$, we define the FOTOC $\mathcal F_G(\tau,\delta\phi)=|\langle\psi_0|\e^{i\delta\phi G(\tau)}|\psi_0\rangle|^2$, with $\tau=\omega t$ \cite{LewisSwan2019}.
Its normalized weak-kick response equals the probe variance,
\begin{equation}
\mathcal C_G(\tau)=\lim_{\delta\phi\to0}
\frac{1-\mathcal F_G(\tau,\delta\phi)}{\delta\phi^2}
=\Var_{\psi_0}[G(\tau)]. \label{eq:fotoc}
\end{equation}

We prepare a wave packet centered at the origin, $\lvert\psi_0\rangle=\lvert0\rangle\otimes\lvert\downarrow\rangle$, where $\lvert0\rangle$ is the field vacuum and $\lvert\downarrow\rangle$ the lower spin state.
Using the momentum quadrature $p=i(\hat a^\dagger-\hat a)/2$ as the probe, we compare coupling paths through the largest increase in its variance during $0\leq\tau\leq1$:
\begin{equation}
S_o=\log_{10}\!\left\{1+
\max_{0\leq\tau\leq1}\left[\mathcal C_p(\tau)-\mathcal C_p(0)\right]\right\}.
\label{eq:origin-response}
\end{equation}
\begin{figure}[t]
\includegraphics[width=\columnwidth]{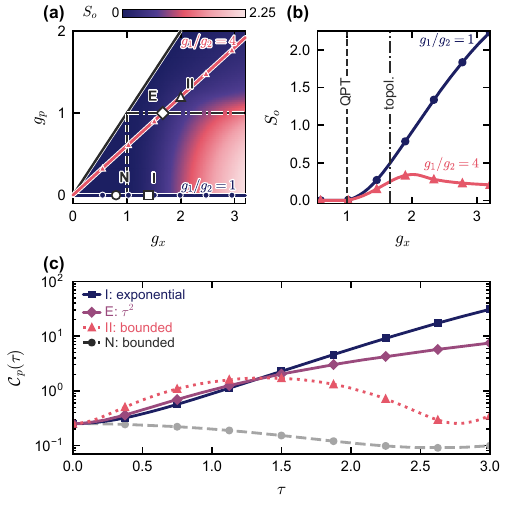}
\caption{\label{fig:rabi-map}ARM scrambling probes the stability of the origin ($\eta=1600$).
(a,b) Origin response $S_o$ across the coupling plane and along the labeled paths, respectively.
The solid diagonal in (a) marks the integrable limit $g_2=0$.
(c) Quantum dynamics of $\mathcal C_p(\tau)$ for origin packets at N, I, E, and II marked in (a) \cite{SupplementalMaterial}.}
\end{figure}

The origin becomes unstable as $g_x$ increases through the QPT boundary $g_x=1$.
Both the $g_1/g_2=1$ and $4$ paths cross this boundary, whereas only the latter crosses $g_p=1$ at $g_x=5/3$, beyond which the origin regains stability.
Accordingly, $S_o$ rises after the QPT along both paths, then eventually decreases for $g_1/g_2=4$ while continuing to increase for $g_1/g_2=1$ [Figs.~\ref{fig:rabi-map}(a) and \ref{fig:rabi-map}(b)].
This difference occurs entirely within the superradiant phase.
The quantum dynamics confirms this stability picture: N and II show bounded responses, I shows exponential growth, and E at $g_p=1$ shows growth proportional to $\tau^2$ [Fig.~\ref{fig:rabi-map}(c)] \cite{SupplementalMaterial}.
The FOTOC therefore tracks the stability of the origin, explaining the apparent connection between scrambling and the QPT.

\begin{figure}[t]
\centering
\includegraphics[width=\columnwidth]{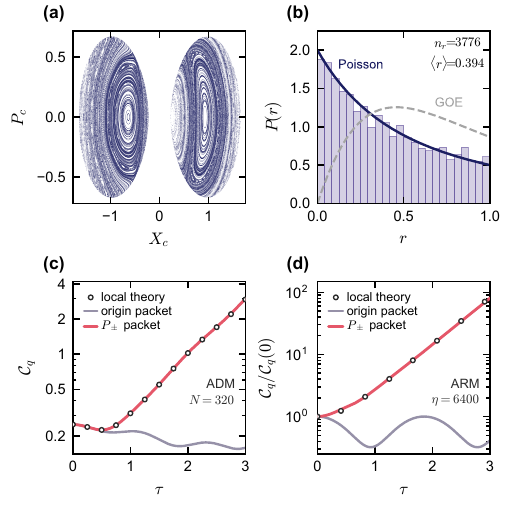}
\caption{\label{fig:sameshell}Global chaos is not required for local exponential FOTOC growth.
(a) ADM Poincar\'e section at $h_P$.
(b) Within-parity adjacent-gap ratios at $N=400$, compared with Poisson and GOE predictions.
(c,d) Quantum responses for origin and $P_\pm$ packets: $\mathcal C_q$ for the ADM ($N=320$) and $\mathcal C_q/\mathcal C_q(0)$ for the ARM ($\eta=6400$).
Origin controls have $\epsilon_O=-1/2$, distinct from the saddle energy $\epsilon_P$.
Open circles give local predictions without fitted rates or amplitudes \cite{SupplementalMaterial}.
All panels use $(g_x,g_p)=(2,1.4)$.}
\end{figure}

\emph{Localized scrambling on a globally regular shell.---}
To test whether local exponential growth requires global chaos, we choose the pair of SP-II saddles $P_\pm$ at $(g_x,g_p)=(2,1.4)$, with saddle energy $h_P$ and reduced energy $\epsilon_P=2h_P\simeq-0.61755$ \cite{SupplementalMaterial}.
On this ADM energy shell, nested invariant curves in the Poincar\'e section and the absence of chaotic trajectories among 4096 Lyapunov samples indicate regular classical dynamics [Fig.~\ref{fig:sameshell}(a)] \cite{SupplementalMaterial}.
The quantum spectrum also supports this picture of global regularity: at $N=400$, the within-parity adjacent-gap ratios follow Poisson statistics, with $\langle r\rangle=0.394$ [Fig.~\ref{fig:sameshell}(b)] \cite{Atas2013,SupplementalMaterial}.

\begin{figure}[t]
\centering
\includegraphics[width=\columnwidth]{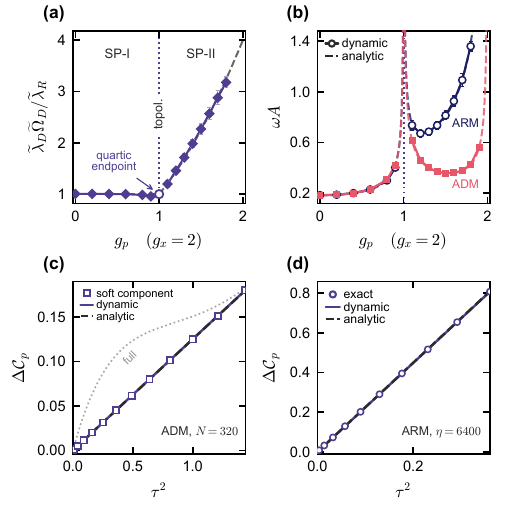}
\caption{\label{fig:spectroscopy}Local dynamics reconstructs the absolute amplitudes of spectral singularities ($g_x=2$).
(a) Dynamical $\tlambda_D\tOmega_D/\tlambda_R$ (diamonds) and Eq.~\eqref{eq:maps} (dashed).
The open circle at $g_p=1$ marks their common quartic limit.
(b) Reconstructed $\omega A_R$ (blue) and $\omega\bar A_D$ (red), with analytic amplitudes dashed \cite{SupplementalMaterial}.
Markers in (a,b) show large-size extrapolations; error bars combine numerical and extrapolation variations.
(c,d) Endpoint ADM and ARM: exact dynamics (markers), reconstructed responses (solid), and analytic predictions (dashed) \cite{SupplementalMaterial}.
Squares in (c) show the response after subtracting the stable oscillation from the full signal (gray dotted).}
\end{figure}

Nevertheless, a wave packet centered at a $P_\pm$ saddle on this ADM shell shows exponential growth of $\mathcal C_q$, with $q=(\hat a^\dagger+\hat a)/2$ [Fig.~\ref{fig:sameshell}(c)].
The local theory (open circles) evolves the initial fluctuations with the quadratic Hamiltonian around the saddle, predicting the probe variance without fitting a rate or amplitude [Eq.~(S5), Supplemental Material \cite{SupplementalMaterial}].
The predictions agree with quantum dynamics in both models at the same reduced saddle energy [Figs.~\ref{fig:sameshell}(c) and \ref{fig:sameshell}(d)].
By contrast, packets centered at the origin remain bounded at the same coupling, at the different reduced energy $\epsilon_O=-1/2$.
This exponential growth therefore originates from the local saddle instability and does not require global chaos.

\emph{Scrambling spectroscopy.---}
Local FOTOC dynamics also provides quantitative spectral information, since the same local Hamiltonian controls both the response and the leading singularity in the density of states (DOS).
For quadratic saddles, the ARM DOS has a logarithmic singularity, whereas integrating over the ADM's additional stable mode places this singularity in the DOS derivative.
Denoting the Rabi and Dicke densities of states by $\nu_R$ and $\nu_D$, we define the positive singular amplitudes near the saddle energy $\epsilon_s$ by \cite{SupplementalMaterial}
\begin{equation}
\begin{gathered}
\nu_R(\epsilon)=\nu_R^{\rm reg}(\epsilon)-A_R\ln|\epsilon-\epsilon_s|+\cdots,\\
\frac{\dd}{\dd\epsilon}\frac{\nu_D(\epsilon)}{N}
={\rm reg}-\bar A_D\ln|\epsilon-\epsilon_s|+\cdots .
\end{gathered}
\label{eq:dosdefs}
\end{equation}

At a quadratic saddle, these amplitudes are set by the local instability rate and, for the ADM, the stable-mode frequency.
Writing $\tlambda_{R,D}=\lambda_{R,D}/\omega$ and $\tOmega_D=\Omega_D/\omega$, integration over $M$ symmetry-related saddles gives \cite{StranskyCejnar2016,Brandes2013}
\begin{equation}
\omega A_R=\frac{M}{\pi\tlambda_R},\qquad
\omega\bar A_D=\frac{M}{\pi\tlambda_D\tOmega_D}.
\label{eq:invariants}
\end{equation}
Topology gives $M=1$ for the SP-I origin and $M=2$ for the SP-II $P_\pm$ pair.
The ARM response yields $\tlambda_R$.
At the same ADM saddle, a packet overlapping the unstable mode yields $\tlambda_D$ from its growing response, whereas a stable-mode squeezed packet yields $\tOmega_D$ from its oscillations.
Thus Eq.~\eqref{eq:invariants} converts dynamical scales into absolute amplitudes without fitting the DOS.

The shared stationary-point geometry predicts when the two models have equal singular amplitudes and when they differ.
At the same $(g_x,g_p)$, expanding the adiabatic ARM and resonant ADM Hamiltonians to quadratic order around the corresponding saddles gives
\begin{equation}
\frac{\tlambda_D\tOmega_D}{\tlambda_R}=
\begin{cases}
1,&\mathrm{SP\mbox{-}I},\\
g_p^2,&\mathrm{SP\mbox{-}II}.
\end{cases}
\label{eq:maps}
\end{equation}
Consequently, the amplitudes coincide in SP-I, whereas the ADM amplitude is $g_p^{-2}$ times the ARM amplitude in SP-II.
At $g_x=2$, the dynamical rate ratio follows Eq.~\eqref{eq:maps} [Fig.~\ref{fig:spectroscopy}(a)], and the reconstructed amplitudes agree with the analytic predictions \cite{SupplementalMaterial} to within $4.5\%$ across the tested couplings [Fig.~\ref{fig:spectroscopy}(b)].
The plotted extrapolations include sequences limited by the crossover near $g_p=1$ \cite{SupplementalMaterial}.

As $g_p\to1$, $\tlambda_D$ and $\tlambda_R$ vanish, while $\tlambda_D\tOmega_D/\tlambda_R$ approaches unity and the quadratic amplitudes diverge [Figs.~\ref{fig:spectroscopy}(a) and \ref{fig:spectroscopy}(b)].
At $g_p=1$, the vanishing quadratic curvature along $P$ requires retaining the quartic term in the DOS calculation:
\begin{equation}
\varepsilon_-+\frac12=-\frac{B_R}{2}X^2+\frac14P^4+\cdots,
\qquad B_R=g_x^2-1. \label{eq:quartic}
\end{equation}
Here $B_R$ is the magnitude of the negative curvature along $X$.
Integrating this quartic normal form gives the same $|\epsilon+1/2|^{-1/4}$ singularity in $\nu_R$ and $\dd(\nu_D/N)/\dd\epsilon$:
\begin{equation}
\begin{gathered}
\nu_R^{\rm sing},\ 
\left[\frac{\dd}{\dd\epsilon}\frac{\nu_D}{N}\right]_{\!\rm sing}
=A_\pm\left|\epsilon+\frac12\right|^{-1/4},\\
A_-=\frac{\mathrm B(1/4,1/4)}{2\pi\omega\sqrt{B_R}},\quad
A_+=\frac{\mathrm B(1/4,1/2)}{2\pi\omega\sqrt{B_R}}.
\end{gathered}
\label{eq:quarticamp}
\end{equation}
Here $A_-$ and $A_+$ apply below and above $\epsilon=-1/2$, respectively; $\mathrm B$ is the Euler beta function, and $A_-/A_+=\sqrt2$ \cite{YuanIsobeFu2019,Hu2023}.

Although exponential growth disappears at $g_p=1$, the remaining algebraic response still encodes the curvature needed for reconstruction.
For the vacuum covariances used here, the ARM variance increase $\Delta\mathcal C_p(\tau)=\mathcal C_p(\tau)-\mathcal C_p(0)$ grows as $\tau^2$; the same law holds for the ADM after subtracting the bounded stable-mode oscillation [Figs.~\ref{fig:spectroscopy}(c) and \ref{fig:spectroscopy}(d)].
The growth coefficient determines $B_R$ from the known initial covariance, with the ADM additionally requiring the stable-mode frequency extracted from the full response \cite{SupplementalMaterial}.
Combined with the known quartic coefficient in Eq.~\eqref{eq:quartic}, this curvature fixes the absolute amplitudes in Eq.~\eqref{eq:quarticamp} without fitting the DOS.
The reconstructed amplitudes agree with their analytic values \cite{SupplementalMaterial}, extending the spectroscopic reconstruction to both sides of the quartic singularity.

\emph{Conclusion.---}
Our comparison shows how scrambling separates superradiance from quantum chaos.
On fixed-energy shells, the anisotropic Dicke model can be chaotic in the normal phase and regular in the superradiant phase, making superradiance neither necessary nor sufficient for global chaos.
An isotropic sweep at fixed low excitation energy crosses both the superradiant boundary and a nearby chaotic window, explaining their reported correlation.

A localized FOTOC instead probes stability at the center of the initial wave packet.
The superradiant transition destabilizes the origin, explaining the apparent connection between scrambling and superradiance.
Anisotropy can restore origin stability while creating displaced saddles within the superradiant phase.
Packets centered at these saddles show exponential FOTOC growth on a classically regular ADM shell with Poisson statistics, demonstrating that global chaos is not required.

This local sensitivity also enables quantitative spectroscopy.
Combined with stationary-point topology and local DOS relations, both exponential and algebraic FOTOC responses allow us to reconstruct absolute amplitudes of spectral singularities in both models without fitting the DOS.

\begin{acknowledgments}
This work was supported by the National Natural Science Foundation of China (Grant No.~12205383) and the Hunan Provincial Natural Science Foundation (Grant No.~2024JJ6483).
\end{acknowledgments}

\end{document}


\title{Supplemental Material for ``Scrambling Separates Superradiance from Quantum Chaos''}

\author{Long-Jie Li}
\affiliation{Institute of Quantum Physics, Hunan Key Laboratory of Nanophotonics and Devices, Hunan Key Laboratory of Super-Microstructure and Ultrafast Process, School of Physics, Central South University, Changsha 410083, China}

\author{Li-Li Gao}
\affiliation{Institute of Quantum Physics, Hunan Key Laboratory of Nanophotonics and Devices, Hunan Key Laboratory of Super-Microstructure and Ultrafast Process, School of Physics, Central South University, Changsha 410083, China}

\author{Zi-Cheng Liu}
\affiliation{Institute of Quantum Physics, Hunan Key Laboratory of Nanophotonics and Devices, Hunan Key Laboratory of Super-Microstructure and Ultrafast Process, School of Physics, Central South University, Changsha 410083, China}

\author{Zi-Min Li}
\email{zimin.li@csu.edu.cn}
\affiliation{Institute of Quantum Physics, Hunan Key Laboratory of Nanophotonics and Devices, Hunan Key Laboratory of Super-Microstructure and Ultrafast Process, School of Physics, Central South University, Changsha 410083, China}

\author{Murray T. Batchelor}
\affiliation{Mathematical Sciences Institute, Australian National University, ACT 2601, Australia}

\date{October 20, 2026}
\maketitle

This Supplemental Material fixes the conventions used in the spectroscopy
relations, derives the local density-of-states (DOS) coefficients and the
general covariance-calibrated endpoint protocol, and documents the controlled
$\eta\to\infty$ and $N\to\infty$ extrapolations.  It also gives the classical
and spectral checks supporting the main text.  All quantitative
ADM calculations use $\Delta=\omega$, whereas the ARM calculations approach
the controlled limit $\eta=\Delta/\omega\to\infty$.  The cross-model
identities compare these limits at the same dimensionless couplings.

\section{Conventions and normalized fidelity loss}
\label{sec:sm-conventions}

The coupling coordinates are
\begin{equation}
\begin{aligned}
g_c&=\sqrt{\Delta\omega}, &
g_x&=\frac{g_1+g_2}{g_c}, &
g_p&=\frac{g_1-g_2}{g_c}.
\end{aligned}
\label{eq:sm-kappa}
\end{equation}
Thus $g_{1,2}=\sqrt{\Delta\omega}(g_x\pm g_p)/2$.
We consider $g_1\geq g_2\geq0$, or $0\leq g_p\leq g_x$, throughout.
We measure time by $\tau=\omega t$.  A small displacement along a locally
unstable mode grows as $e^{\tlambda\tau}$, whereas a stable mode oscillates as
$e^{\pm i\tOmega\tau}$; these relations define the dimensionless local
instability rate $\tlambda$ and stable-mode frequency $\tOmega$.
The quadratures
$q=(a+a^\dagger)/2$ and $p=i(a^\dagger-a)/2$ obey $[q,p]=i/2$ and have
vacuum covariance $\Var q=\Var p=1/4$.
In the controlled ARM limit, $X=\sqrt{2/\eta}\,q$ and
$P=\sqrt{2/\eta}\,p$; for the resonant ADM,
$X=\sqrt{2/N}\,q$ and $P=\sqrt{2/N}\,p$.
The corresponding reduced lower-branch energies are $\epsilon=E/\Delta$ for
the ARM and $\epsilon=E/(N\Delta)=2h$ for the ADM.  The
$O(\eta^{-1})$ oscillator zero-point offset is omitted from the limiting ARM
surface.

To avoid calling a fidelity loss a fidelity, let
\begin{equation}
\mathcal F_G(\tau,\delta\phi)
=\left|\langle\psi_0|
\e^{i\delta\phi G(\tau)}|\psi_0\rangle\right|^2
\label{eq:sm-fidelity}
\end{equation}
and define the normalized weak-kick response
\begin{equation}
\mathcal C_G(\tau)
=\lim_{\delta\phi\rightarrow0}
\frac{1-\mathcal F_G(\tau,\delta\phi)}{\delta\phi^2}
=\Var_{\psi_0}[G(\tau)].
\label{eq:sm-response}
\end{equation}
For local fluctuation coordinates $\bm z$, write
$\Delta z_i=z_i-\langle z_i\rangle_{\psi_0}$.  The initial symmetrized covariance is
\begin{equation}
(\Sigma_0)_{ij}=\frac12\langle\{\Delta z_i,\Delta z_j\}\rangle_{\psi_0}.
\label{eq:sm-initial-covariance}
\end{equation}
Linearizing the dynamics around a stationary point $s$ gives
$\dot{\bm z}=\mathsf L_s\bm z$, where the derivative is with respect to $\tau$.
The local fluctuations therefore evolve as
$\bm z(\tau)=M_s(\tau)\bm z(0)$, with $M_s(\tau)=\e^{\mathsf L_s\tau}$.
For a linear probe $G=\bm g^{\mathsf T}\bm z$, the coefficient vector $\bm g$
selects the measured combination, and the local response is
\begin{equation}
\mathcal C_G^{\rm loc}(\tau)
=\bm g^{\mathsf T}M_s(\tau)\Sigma_0M_s^{\mathsf T}(\tau)\bm g.
\label{eq:sm-local-covariance}
\end{equation}
For the adiabatic ARM, $\bm z=(\delta q,\delta p)^{\mathsf T}$ contains
deviations from the stationary values; at the origin, $G=p$ corresponds to
$\bm g=(0,1)^{\mathsf T}$.
The full local ADM dynamics also includes spin fluctuations.
Equation~\eqref{eq:sm-local-covariance} supplies the parameter-free predictions
in Figs.~4(c) and 4(d) of the main text: the Hamiltonian fixes $M_s$, the
prepared state fixes $\Sigma_0$, and the probe fixes $\bm g$.

For an isolated unstable canonical pair with linear generator $\mathsf L$,
write $\mathsf K=\mathsf L/\tlambda$, with $\mathsf K^2=I$.
The propagator is then
$M(\tau)=I\cosh(\tlambda\tau)+\mathsf K\sinh(\tlambda\tau)$.
For $G=\bm g^{\mathsf T}\bm z$, define the variance increase
$\Delta\mathcal C_G(\tau)=\mathcal C_G(\tau)-\mathcal C_G(0)$.
Substitution into Eq.~\eqref{eq:sm-local-covariance} gives
\begin{equation}
\begin{aligned}
\Delta\mathcal C_G(\tau)
&={}C_1\sinh(\tlambda\tau)\cosh(\tlambda\tau)\\
&+C_2\sinh^2(\tlambda\tau),\\
C_1&={}\bm g^{\mathsf T}(\mathsf K\Sigma_0+\Sigma_0\mathsf K^{\mathsf T})\bm g,\\
C_2&={}\bm g^{\mathsf T}(\Sigma_0+\mathsf K\Sigma_0\mathsf K^{\mathsf T})\bm g.
\end{aligned}
\label{eq:sm-saddle-response}
\end{equation}
Thus the pure $\sinh^2$ response requires $C_1=0$.
For the aligned ARM packets at the origin and $P_\pm$ saddles,
$\mathsf L$ has zero diagonal entries in the local $q,p$ coordinates and
$\Sigma_{qp}=0$.  Measuring either $q$ or $p$ therefore gives $C_1=0$,
with the amplitude fixed by $C_2$.
The slope of $\ln\Delta\mathcal C_G$ in the exponential part of the local
window is $2\tlambda$.
The ADM cavity response also contains the stable mode and is evaluated with
the full covariance propagator.

The weak-kick and thermodynamic limits must be taken in that order.  For a
Gaussian quadrature distribution, a finite kick gives
\begin{equation}
\begin{aligned}
\mathcal C_{G,\delta\phi}
&=\frac{1-\exp[-\delta\phi^2\Var G]}{\delta\phi^2}\\
&=\Var G+O(\delta\phi^2\Var^2G).
\end{aligned}
\label{eq:sm-finite-kick}
\end{equation}
Because the fluctuation variance grows with the semiclassical size, a fixed
kick can saturate even when the covariance response is well converged.  At
$g_x=2$, $g_p=1.5$, and $\eta=25600$, the kick $\delta\phi=0.1$ no longer
resolves the response interval used for rate extraction, whereas $0.05$ and
$0.025$ do.  A
zero-kick extrapolation from the finite-kick traces differs from the directly
calculated covariance rate by $0.0603\%$; omitting the largest kick changes
that extrapolation by $0.0477\%$.  We therefore use Eq.~\eqref{eq:sm-response}
for the thermodynamic extrapolation and treat fixed-kick saturation as the
expected nonuniformity of Eq.~\eqref{eq:sm-finite-kick}, not as a failure of
the weak-kick response.

We express the DOS per unit physical energy as a function of reduced energy:
\begin{align}
\nu_R(\epsilon)
&=\frac{1}{2\pi\omega}\int\dd X\,\dd P\,
\delta[\epsilon-\varepsilon_-(X,P)]+\nu_R^{(+)}(\epsilon),
\label{eq:sm-rabi-dos}\\
\frac{\nu_D(\epsilon)}{N}
&=\frac{1}{2\pi^2\omega}\int\dd^4z\,
\delta\!\left(\frac{\epsilon}{2}-h(\bm z)\right).
\label{eq:sm-dicke-dos}
\end{align}
These prefactors follow from the semiclassical trace measure.
Because $[q,p]=i/2$, the oscillator trace uses the measure $\dd q\,\dd p/\pi$.
The transformations to $X,P$ contribute Jacobians $\eta/2$ and $N/2$ in the ARM and ADM, respectively.
Combining these Jacobians with $E=\eta\omega\epsilon$ for the ARM, $E=N\omega\epsilon$ for the resonant ADM, and the normalized spin-sphere measure gives the factors $1/(2\pi\omega)$ and $1/(2\pi^2\omega)$ above.
Here $\epsilon=2h$ in the ADM convention.
The second term in Eq.~\eqref{eq:sm-rabi-dos} denotes the nonsingular upper
adiabatic branch in the energy windows used below.
In Eq.~\eqref{eq:sm-dicke-dos}, $\bm z$ denotes canonical ADM phase-space
coordinates and $\dd^4z$ their Liouville measure.  Near a quadratic saddle
we use $\bm z=(Q_s,P_s,Q_u,P_u)$; at the quartic endpoint the pairs are
denoted by $(Q_+,P_+)$ and $(X_-,P_-)$.

\section{Quadratic scrambling spectroscopy}
\label{sec:sm-morse}

This section derives the analytic curves in Figs.~5(a) and 5(b) of the main
text.  The calculation combines four ingredients: the local expansion fixes
the curvature dependence, stationary-point multiplicity fixes the prefactor,
finite-size time evolution reconstructs the local dynamical scales, and
phase-space integration supplies an independent numerical value of each DOS
coefficient.

\subsection{Local DOS invariants}

The lower Rabi surface is
\begin{equation}
\varepsilon_-(X,P)=\frac{X^2+P^2}{2}
-\frac12\sqrt{1+2g_x^2X^2+2g_p^2P^2}.
\label{eq:sm-surface}
\end{equation}
It follows from the controlled limit $\eta=\Delta/\omega\to\infty$ after
adiabatically selecting the lower spin branch.  The resulting ARM surface
has one canonical pair $(X,P)$.

Near an index-one ARM saddle, a local canonical transformation gives
\begin{equation}
\varepsilon_--\epsilon_s
=\frac{\tlambda_R}{2}(v^2-u^2)+O(3).
\label{eq:sm-rabi-normal}
\end{equation}
With $\delta\epsilon=\epsilon-\epsilon_s$, the singular part of its shell integral is
\begin{equation}
\begin{aligned}
I_R(\delta\epsilon)
&=\int\dd u\,\dd v\,
\delta\!\left[\delta\epsilon
-\frac{\tlambda_R}{2}(v^2-u^2)\right]\\
&=-\frac{2}{\tlambda_R}\ln|\delta\epsilon|+\mathrm{reg}.
\end{aligned}
\label{eq:sm-rabi-shell-integral}
\end{equation}
Here $\mathrm{reg}$ denotes terms regular at the saddle energy.
Including the DOS normalization in Eq.~\eqref{eq:sm-rabi-dos} and $M$
symmetry-related saddles gives
\begin{equation}
\begin{aligned}
\nu_R(\epsilon)
&=\nu_R^{\rm reg}(\epsilon)
-A_R\ln|\epsilon-\epsilon_s|+\cdots,\\
(\omega A_R)\tlambda_R&=\frac{M}{\pi}.
\end{aligned}
\label{eq:sm-rabi-invariant}
\end{equation}

The ADM retains two canonical pairs.  At the relevant saddle, its local
normal form contains one stable and one hyperbolic mode,
\begin{equation}
h-h_s=\frac{\tOmega_D}{2}(P_s^2+Q_s^2)
+\frac{\tlambda_D}{2}(P_u^2-Q_u^2)+O(3).
\label{eq:sm-dicke-normal}
\end{equation}
The stable polar integral contributes $1/\tOmega_D$; differentiating the
remaining shell integral produces the logarithm.  With the normalization in
Eq.~\eqref{eq:sm-dicke-dos},
\begin{equation}
\begin{aligned}
\frac{\dd}{\dd\epsilon}\frac{\nu_D(\epsilon)}{N}
&=\mathrm{reg}-\bar A_D\ln|\epsilon-\epsilon_s|+\cdots,\\
(\omega\bar A_D)\tlambda_D\tOmega_D&=\frac{M}{\pi}.
\end{aligned}
\label{eq:sm-dicke-invariant}
\end{equation}
If $\mathsf L_D$ is the linearized ADM Hamiltonian flow, its saddle--center
spectrum is $\{\pm\tlambda_D,\pm i\tOmega_D\}$.  The basis-independent
local scale entering the DOS coefficient is therefore
\begin{equation}
K_D\equiv\tlambda_D\tOmega_D=\sqrt{-\det\mathsf L_D}.
\label{eq:sm-kd}
\end{equation}
The shorthand $K_D$ used here denotes the product $\tlambda_D\tOmega_D$
written explicitly in the main text and Fig.~5(a).
Equations~\eqref{eq:sm-rabi-invariant} and
\eqref{eq:sm-dicke-invariant} then read
\begin{equation}
\begin{aligned}
\omega A_R&=\frac{M}{\pi\tlambda_R}, &
\omega\bar A_D&=\frac{M}{\pi K_D}.
\end{aligned}
\label{eq:sm-dynamic-amplitudes}
\end{equation}
The ADM amplitude depends on both local modes because its energy shell has
one more canonical pair than the ARM surface.

The local FOTOC response determines the dynamical scales of one saddle,
whereas the DOS adds the contributions of all symmetry-related saddles.
The multiplicity $M$ must therefore be supplied independently by the
stationary-point topology.  With this input, Eq.~\eqref{eq:sm-dynamic-amplitudes}
reconstructs the absolute amplitudes without fitting the DOS or an
intermodel conversion factor.

\subsection{Stationary-point curvatures and cross-model map}

At the SP-I origin, Eq.~\eqref{eq:sm-surface} expands as
\begin{equation}
\varepsilon_-+\frac12
=\frac{1-g_x^2}{2}X^2+\frac{1-g_p^2}{2}P^2+O(4),
\label{eq:sm-rabi-origin-expansion}
\end{equation}
which gives
\begin{equation}
\tlambda_R=\sqrt{(g_x^2-1)(1-g_p^2)}.
\label{eq:sm-rabi-origin-rate}
\end{equation}
For the resonant ADM, $(q_a,p_a)=(q,p)$ are the cavity quadratures.
The Holstein--Primakoff boson $b$ describes spin fluctuations about the
south pole, with $J_z=b^\dagger b-N/2$ and $J_-\simeq\sqrt N\,b$.
Its quadratures are $q_b=(b+b^\dagger)/2$ and
$p_b=i(b^\dagger-b)/2$.
With
$q_\pm=(q_a\pm q_b)/\sqrt2$ and
$p_\pm=(p_a\pm p_b)/\sqrt2$, the quadratic fluctuation Hamiltonian at the
same origin is
\begin{equation}
\begin{aligned}
\frac{H_D^{(2)}}{\omega}&={}
(1+g_x)q_+^2+(1+g_p)p_+^2\\
&+(1-g_x)q_-^2+(1-g_p)p_-^2.
\end{aligned}
\label{eq:sm-dicke-origin-quadratic}
\end{equation}
The minus mode is hyperbolic and the plus mode is stable in SP-I, giving
\begin{equation}
\begin{aligned}
\tlambda_D&=\sqrt{(g_x-1)(1-g_p)},\\
\tOmega_D&=\sqrt{(g_x+1)(1+g_p)},\\
K_D&=\tlambda_R.
\end{aligned}
\label{eq:sm-origin-map}
\end{equation}
The origin is a single saddle; hence $M=1$.

In SP-II, Eq.~\eqref{eq:sm-surface} has two saddles at
\begin{equation}
\begin{aligned}
P_\pm^2&=\frac{g_p^4-1}{2g_p^2}, &
\epsilon_P&=-\frac14\left(g_p^2+g_p^{-2}\right).
\end{aligned}
\label{eq:sm-rabi-p-saddles}
\end{equation}
Writing $\delta P=P-P_\pm$, the quadratic expansion is
\begin{equation}
\begin{aligned}
\varepsilon_--\epsilon_P
&=\frac12\left(1-\frac{g_x^2}{g_p^2}\right)X^2\\
&\quad+\frac12\left(1-g_p^{-4}\right)\delta P^2+O(3),
\end{aligned}
\label{eq:sm-rabi-p-expansion}
\end{equation}
and hence
\begin{equation}
\tlambda_R^{(P)}=
\sqrt{\left(\frac{g_x^2}{g_p^2}-1\right)
\left(1-g_p^{-4}\right)}.
\label{eq:sm-rabi-p-rate}
\end{equation}
For the ADM, write $\bm s=2\langle\bm J\rangle/N$ and
$(X_c,P_c)=(X,P)/\sqrt2$.
The corresponding saddles have
\begin{equation}
\begin{aligned}
(s_x,s_y,s_z)&=
\left(0,\ \pm\sqrt{1-g_p^{-4}},\ -g_p^{-2}\right),\\
(X_c,P_c)&=
\left(0,\ \pm\frac12\sqrt{g_p^2-g_p^{-2}}\right).
\end{aligned}
\label{eq:sm-dicke-p-saddles}
\end{equation}
The constrained ADM linearization at either saddle is specified by
\begin{equation}
\begin{aligned}
\mathcal T_P&=g_p^4+1+\frac{2g_x}{g_p},\\
\mathcal Q_P&=\frac{(g_p^4-1)(g_x^2-g_p^2)}{g_p^2}.
\end{aligned}
\label{eq:sm-dicke-p-invariants}
\end{equation}
Its two local scales are
\begin{equation}
\begin{aligned}
\bigl(\tlambda_D^{(P)}\bigr)^2
&=\frac{\sqrt{\mathcal T_P^2+4\mathcal Q_P}-\mathcal T_P}{2},\\
\bigl(\tOmega_D^{(P)}\bigr)^2
&=\frac{\sqrt{\mathcal T_P^2+4\mathcal Q_P}+\mathcal T_P}{2}.
\end{aligned}
\label{eq:sm-dicke-p-rates}
\end{equation}
Their product is fixed without evaluating either square root separately,
\begin{equation}
K_D^{(P)}=\sqrt{\mathcal Q_P}
=g_p^2\tlambda_R^{(P)}.
\label{eq:sm-p-map}
\end{equation}
The two symmetry-related $P_\pm$ saddles give $M=2$.  Combining the two
topologies yields the analytic map plotted in Fig.~5(a),
\begin{equation}
\frac{K_D}{\tlambda_R}=
\begin{cases}
1, & \mathrm{SP\mbox{-}I},\\
g_p^2, & \mathrm{SP\mbox{-}II},
\end{cases}
\label{eq:sm-cross-model-map}
\end{equation}
and the absolute amplitudes plotted as dashed curves in Fig.~5(b),
\begin{equation}
\begin{aligned}
&(\omega A_R,\omega\bar A_D)\\
&\quad=\begin{cases}
\displaystyle
\frac{(1,1)}{\pi\sqrt{(g_x^2-1)(1-g_p^2)}},
& \mathrm{SP\mbox{-}I},\\[3mm]
\displaystyle
\frac{2(1,g_p^{-2})}
{\pi\sqrt{(g_x^2/g_p^2-1)(1-g_p^{-4})}},
& \mathrm{SP\mbox{-}II}.
\end{cases}
\end{aligned}
\label{eq:sm-analytic-amplitudes}
\end{equation}
Thus the two amplitudes coincide in SP-I and satisfy
$\bar A_D^{(P)}/A_R^{(P)}=g_p^{-2}$ in SP-II.

\subsection{Independent phase-space integration}

The dashed curves in Fig.~5(b) are evaluated from
Eq.~\eqref{eq:sm-analytic-amplitudes}.  Independently, we evaluate the phase-
space integrals in Eqs.~\eqref{eq:sm-rabi-dos} and
\eqref{eq:sm-dicke-dos} on 31 logarithmically spaced energies per side.  The
primary window is $10^{-6}\leq|\epsilon-\epsilon_s|\leq10^{-3}$; near the
coalescing origin and $P_\pm$ singularities, its upper edge is reduced to
$0.02|\epsilon_P+1/2|$.  A simultaneous two-sided fit shares one singular
coefficient while assigning an independent second-order background to each
side.  We test window and background sensitivity by dividing the upper edge
by three, multiplying the lower edge by three, and lowering the background
order.  No dynamical rate or intermodel scale enters this calculation.

The audit contains 72 two-sided fits: 36 saddle cases in each of the two
models.  The maximum absolute relative coefficient error is
$3.60\times10^{-5}$ and the median is $8.78\times10^{-8}$.  The same protocol
provides the independent closure entries in
Table~\ref{tab:sm-quadratic-closure}.

\begin{figure}[t]
\includegraphics[width=\linewidth]{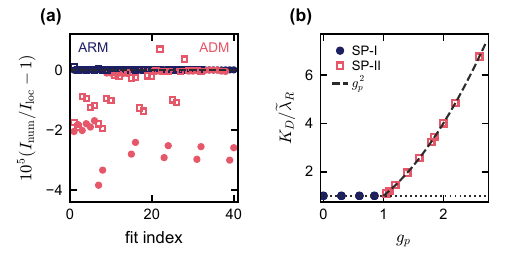}
\caption{Independent phase-space audit of the quadratic singularities.
(a) Residuals of the normalized local invariants; colors distinguish the
ARM and ADM, while circles and open squares denote SP-I and SP-II.
(b) Cross-model local-scale map.  SP-I follows unity (dotted) and SP-II
follows $g_p^2$ (dashed).}
\label{fig:sm-static-morse}
\end{figure}

\section{Quartic endpoint spectroscopy}
\label{sec:sm-endpoint}

The quadratic-saddle rates vanish at $g_p=1$; the endpoint consequently
requires its own local form and dynamical observable.  This section derives both the
$|\delta|^{-1/4}$ singularity and its reconstruction from an algebraic FOTOC response.

\subsection{Quartic DOS integral}

At $g_p=1$, the common weighted local surface is
\begin{equation}
\begin{aligned}
\varepsilon_-+\frac12
&=-\frac{B_R}{2}X^2+\frac14P^4+\cdots,\\
B_R&=g_x^2-1.
\end{aligned}
\label{eq:sm-quartic}
\end{equation}
Let $\delta=\epsilon+1/2$ and let $\delta_{\rm D}$ denote the Dirac delta.
The singular shell integral is
\begin{equation}
\begin{aligned}
I_\pm(\delta)
&=\int\dd X\,\dd P\,
\delta_{\rm D}\!\left(\delta+\frac{B_R}{2}X^2-\frac14P^4\right)\\
&=\frac{\mathcal B_\pm}{\sqrt{B_R}}|\delta|^{-1/4},
\end{aligned}
\label{eq:sm-quartic-integral}
\end{equation}
where the lower and upper sides of the endpoint give
\begin{equation}
\begin{aligned}
\mathcal B_-&=\mathrm B(1/4,1/4), &
\mathcal B_+&=\mathrm B(1/4,1/2).
\end{aligned}
\label{eq:sm-quartic-beta}
\end{equation}
Here $\mathrm B$ is the beta function.
For the ADM, the soft coordinates are $X_-=q_-/\sqrt N$ and
$P_-=p_-/\sqrt N$ to leading order.
With $\tOmega_+=\sqrt{2(g_x+1)}$, the stable coordinates are
$Q_+=\sqrt{(g_x+1)/(N\tOmega_+)}\,q_+$ and
$P_+=\sqrt{2/(N\tOmega_+)}\,p_+$.
The local endpoint form is
\begin{equation}
\begin{aligned}
h+\frac14&={}-\frac{B_D}{2}X_-^2+\frac18P_-^4\\
&+\frac{\tOmega_+}{2}(Q_+^2+P_+^2)+\cdots,\\
B_D&={}g_x-1.
\end{aligned}
\label{eq:sm-dicke-quartic}
\end{equation}
Integration over the stable pair and differentiation with respect to
$\epsilon=2h$ give
\begin{equation}
\begin{aligned}
&\left[\frac{\dd}{\dd\epsilon}\frac{\nu_D}{N}\right]_{\rm sing}\\
&\quad=\frac{1}{\pi\omega\tOmega_+}\int\dd X_-\,\dd P_-\\
&\qquad\times\delta_{\rm D}\!\left(\delta+B_DX_-^2-\frac14P_-^4\right)\\
&\quad=\frac{I_\pm(\delta)}{2\pi\omega}.
\end{aligned}
\label{eq:sm-dicke-quartic-reduction}
\end{equation}
The last equality uses $X=2X_-/\tOmega_+$, $P=P_-$, and
$B_R=\tOmega_+^2B_D/2=g_x^2-1$.  Thus the stable-mode factor, the
$\epsilon=2h$ chain factor, and the different soft curvature reduce exactly
to the Rabi integral.  The phase-space normalizations in
Eqs.~\eqref{eq:sm-rabi-dos} and \eqref{eq:sm-dicke-dos} then give
\begin{equation}
\begin{gathered}
\nu_R^{\rm sing},\quad
\left[\frac{\dd}{\dd\epsilon}\frac{\nu_D}{N}\right]_{\rm sing}
=A_\pm|\delta|^{-1/4},\\
A_\pm=\frac{\mathcal B_\pm}{2\pi\omega\sqrt{B_R}},\qquad
\frac{A_-}{A_+}=\sqrt2.
\end{gathered}
\label{eq:sm-endpoint-amplitudes}
\end{equation}

The amplitude also depends on the quartic coefficient of the local energy
surface.  For $-B X^2/2+c_4P^4/4$, with $B,c_4>0$, the substitutions
$u=\sqrt{B}X$ and $v=c_4^{1/4}P$ give the measure
$\dd X\,\dd P=B^{-1/2}c_4^{-1/4}\dd u\,\dd v$.
With the same DOS normalization, the amplitudes are therefore
\begin{equation}
A_\pm(B,c_4)=\frac{\mathcal B_\pm}{2\pi\omega\sqrt B\,c_4^{1/4}}.
\label{eq:sm-general-quartic-amplitude}
\end{equation}
The linearized endpoint dynamics reads out $B$, while the Hamiltonian
expansion in Eq.~\eqref{eq:sm-quartic} supplies $c_4=1$.
These two inputs give Eq.~\eqref{eq:sm-endpoint-amplitudes}; the ratio
$A_-/A_+=\sqrt2$ is independent of both coefficients.

\subsection{Algebraic dynamical readout}

At the endpoint, the soft-mode generator satisfies $\mathsf L_0^2=0$, and
$M_0(\tau)=I+\tau\mathsf L_0$.
Substitution into Eq.~\eqref{eq:sm-local-covariance} gives
\begin{equation}
\begin{aligned}
\Delta\mathcal C_G(\tau)
&={}\tau\bm g^{\mathsf T}
(\mathsf L_0\Sigma_0+\Sigma_0\mathsf L_0^{\mathsf T})\bm g\\
&+\tau^2\bm g^{\mathsf T}
\mathsf L_0\Sigma_0\mathsf L_0^{\mathsf T}\bm g.
\end{aligned}
\label{eq:sm-general-endpoint}
\end{equation}
The measured initial covariance fixes both coefficients.  A rotated squeezed
packet can generate the linear term, whereas the natural packets used in
Fig.~5(c,d) have no such term.

Below, $\Sigma_{ab}$ denotes the initial symmetrized covariance of
quadratures $a,b$, as in Eq.~\eqref{eq:sm-initial-covariance}.
For the unscaled ARM fluctuations, the endpoint shear is
\begin{equation}
\begin{aligned}
q(\tau)&=q_0, &
p(\tau)&=p_0+B_R\tau q_0.
\end{aligned}
\label{eq:sm-rabi-shear}
\end{equation}
Consequently,
\begin{equation}
\begin{gathered}
\Delta\mathcal C_p(\tau)
=2B_R\Sigma_{qp}\tau+D_R\tau^2,\\
D_R=B_R^2\Sigma_{qq},\qquad
B_R^{\rm dyn}=\sqrt{\frac{D_R}{\Sigma_{qq}}}.
\end{gathered}
\label{eq:sm-rabi-covariance}
\end{equation}

For the resonant ADM, the minus mode shears with $B_D=g_x-1$, while the plus
mode oscillates at $\tOmega_+=\sqrt{2(g_x+1)}$.  The measured cavity
quadrature contains both modes.  For initially uncorrelated modes,
subtracting the bounded plus-mode oscillation leaves the soft component
\begin{equation}
\begin{aligned}
\Delta\mathcal C_{p_a}^{\rm soft}(\tau)
&=B_D\Sigma_{q_-p_-}\tau+D_D\tau^2,\\
D_D&=\frac{B_D^2}{2}\Sigma_{q_-q_-}.
\end{aligned}
\label{eq:sm-dicke-covariance}
\end{equation}
The subtraction in Fig.~5(c) isolates this zero-mode contribution from the
full cavity response while retaining its absolute covariance normalization.
The curvature of the common quartic surface follows from
\begin{equation}
\begin{aligned}
B_D^{\rm dyn}&=\sqrt{\frac{2D_D}{\Sigma_{q_-q_-}}},\\
B_R^{\rm dyn}&=\frac{\tOmega_+^2}{2}B_D^{\rm dyn}.
\end{aligned}
\label{eq:sm-dicke-endpoint-readout}
\end{equation}
The endpoint amplitudes reconstructed from either model are then
\begin{equation}
A_\pm^{\rm dyn}
=\frac{\mathcal B_\pm}{2\pi\omega\sqrt{B_R^{\rm dyn}}}.
\label{eq:sm-endpoint-dynamic-amplitudes}
\end{equation}

For the natural ARM coherent packet and ADM two-mode vacuum,
$\Sigma_{qq}=\Sigma_{q_-q_-}=1/4$ and the cross covariances vanish.  The
analytic coefficients used in Fig.~5(c,d) are therefore
\begin{equation}
\begin{aligned}
D_D&=\frac{(g_x-1)^2}{8}, &
D_R&=\frac{(g_x^2-1)^2}{4}.
\end{aligned}
\label{eq:sm-endpoint-natural-growth}
\end{equation}
Here $D_R$ and $D_D$ are the coefficients of the variance increase
$\Delta\mathcal C_p(\tau)=\mathcal C_p(\tau)-\mathcal C_p(0)$ and its soft component, respectively.
Together with the readout in Eqs.~\eqref{eq:sm-rabi-covariance} and
\eqref{eq:sm-dicke-endpoint-readout}, these coefficients give the absolute
amplitudes in Eq.~\eqref{eq:sm-endpoint-dynamic-amplitudes}.

\begin{figure}[htbp]
\includegraphics[width=\linewidth]{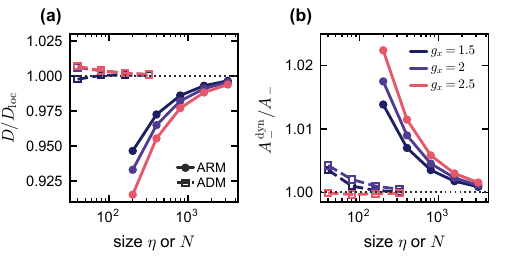}
\caption{Finite-size convergence at $g_p=1$.
(a) Endpoint quadratic coefficient and (b) reconstructed lower-side
amplitude, each normalized by its local analytic value.  Colors give
$g_x=1.5,2,2.5$; filled circles with solid lines and open squares with
dashed lines denote the ARM and ADM.}
\label{fig:sm-endpoint-convergence}
\end{figure}

The convergence scan in Fig.~\ref{fig:sm-endpoint-convergence} uses
$g_x=1.5,2,2.5$, $\eta\leq3200$, and $N\leq320$.
At the largest sizes in that scan, the maximum error in $D$ is $0.618\%$ and
the maximum error in the inferred amplitude is $0.155\%$.  The endpoint
panels at $g_x=2$ use $N=320$ for the ADM and $\eta=6400$ with
$n_{\max}=2048$ for the ARM.  Their dynamical-to-analytic amplitude ratios
are $1.0004907$ and $1.0005765$, respectively.

\section{Exact quantum calculations and truncations}
\label{sec:sm-quantum-method}

For the origin-centered Rabi packet, parity reduces the exact Hamiltonian to
a tridiagonal chain labelled by photon number.  The initial state is the
zero-photon lower-spin state in the appropriate parity sector.  For a
$P$-saddle packet we instead use the full spin--boson Hamiltonian in a
displaced Fock basis.  The displacement equals the macroscopic saddle
coordinate; hence the cutoff controls fluctuations about the saddle rather than
the total coherent occupation.

The Dicke calculation is performed in the exact maximal-spin sector
$j=N/2$, resolved by parity.  It retains the finite spin dimension and does
not use a Holstein--Primakoff truncation for the time evolution.  Sparse
Krylov propagation evaluates the state at a uniform time grid.  For all
calculations the reported maximum norm loss and the probability in the
highest retained basis states are recorded.

The custom Rabi and Dicke parity matrices were independently projected from
the corresponding full Hamiltonians at small dimension.  The maximum matrix
differences are $1.78\times10^{-15}$ and $8.88\times10^{-16}$,
respectively.

\section{Mode-resolved dynamics and asymptotic limits}
\label{sec:sm-dynamics}

The analytic relations above become a spectroscopy protocol only after the
local scales are obtained from the exact finite-size dynamics.  For an ARM
quadratic saddle, we fit the localized response to
\begin{equation}
\Delta\mathcal C_G(\tau)=c\sinh^2(\tlambda_R\tau)
\label{eq:sm-rabi-fit}
\end{equation}
with free amplitude $c$, over the interval in which the packet remains localized.  The beginning and
end of the fit are set by response thresholds, and their variation gives the
window uncertainty.  Paired displaced-basis cutoffs monitor truncation.  The
analytic rate is evaluated only after the numerical sequence passes these
checks.  In the exponential part of the window,
$\ln\Delta\mathcal C_G$ has slope $2\tlambda_R$; the rate entering the DOS
invariant is $\tlambda_R$, rather than this doubled logarithmic slope.

The Dicke saddle has one hyperbolic and one stable local normal mode.  We use
two preparations centered on the same stationary point.  A packet with
finite projection on the hyperbolic mode gives $\tlambda_D$ from its growing
covariance.  A stable-mode squeezed packet gives $\tOmega_D$ from the rotation
of its covariance ellipse.  Their product $K_D$, defined in
Eq.~\eqref{eq:sm-kd}, is the local scale entering
Eq.~\eqref{eq:sm-dicke-invariant}.  The normal-mode coordinates define the
preparations and measured quadratures, while the evolution uses the exact
finite-$N$ Hamiltonian in the maximal-spin sector.

Let $\tlambda_R^{(\infty)}$ and $K_D^{(\infty)}$ denote the controlled
large-$\eta$ and large-$N$ intercepts.  Inserting these intercepts into
Eq.~\eqref{eq:sm-dynamic-amplitudes} gives the dynamical amplitudes in
Fig.~5(b), with the saddle multiplicity fixed by the stationary-point
topology.  To first order, each reconstructed amplitude has the same
relative uncertainty as the corresponding extrapolated local scale.

\begin{figure}[htbp]
\includegraphics[width=\linewidth]{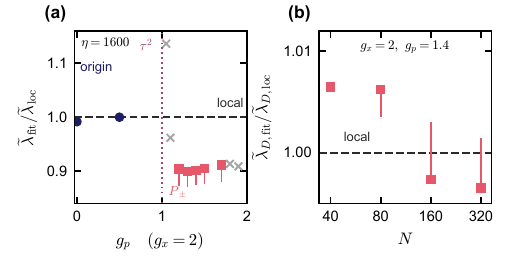}
\caption{Rate-extraction audits.
(a) Rabi fitted-to-local rate ratio at $g_x=2$ and $\eta=1600$.
Blue circles and red squares satisfy the numerical criteria for fits about
the origin and $P_\pm$ saddles; gray crosses do not.
The dotted line marks $g_p=1$, and red vertical segments span three fixed
response windows.
(b) ADM fitted-to-local rate ratio at $(g_x,g_p)=(2,1.4)$ for
$N=40,80,160,320$; vertical segments span the fixed response windows.
Dashed horizontal lines mark the local predictions and are not fits.}
\label{fig:sm-rate-audits}
\end{figure}

Figure~\ref{fig:sm-rate-audits} compares the fitted rates with local predictions.
At fixed $\eta=1600$, fits satisfying the numerical
criteria select the origin saddle below $g_p=1$ and the displaced $P_\pm$
saddles above it.  The
remaining offset on the displaced side is resolved by the large-$\eta$
sequences below.  At $(g_x,g_p)=(2,1.4)$, the finite-$N$ Dicke rates remain
within about $1\%$ of the local value.  The quartic point itself has algebraic
rather than exponential growth and therefore carries no fitted rate.

Tables~\ref{tab:sm-rabi-asymptotic} and \ref{tab:sm-dicke-asymptotic} summarize all
29 finite-size sequences used for the quadratic-saddle spectroscopy in
Fig.~5 of the main text.  Parity makes $1/\eta$ the leading correction at the
Rabi origin; expansion about a displaced saddle permits a leading
$1/\sqrt\eta$ correction, while the Dicke Holstein--Primakoff expansion gives
$1/N$.  Each listed limit is the intercept of a linear fit in the corresponding
variable.  Its uncertainty is the largest change from the fit standard
error, removing the smallest size, adding a quadratic correction, or changing
the response window and cutoff.  For displaced Rabi saddles, it also includes
the shift from a linear $1/\eta$ fit, which is at most $2.89\%$.  The numerical
criteria use only response-window, cutoff, norm, and extrapolation stability;
agreement with the analytic curves is evaluated afterward.

\begin{table*}[t]
\caption{Complete Rabi sequences at $g_x=2$.  The extracted values and limits
are $\tlambda_R$, and $O$ denotes the origin.  The status column marks asymptotically controlled (C) and
crossover-limited (X) extrapolations according to the criteria in
Sec.~\ref{sec:sm-crossover}.}
\label{tab:sm-rabi-asymptotic}
\squeezetable
\scriptsize
\begin{ruledtabular}
\begin{tabular}{lcccccc}
Saddle & $g_p$ & $\eta$ & extracted values & variable & limit & status\\
\hline
$O$ & $0$ & $1600,3200,6400,12800$ & $1.710628,1.715956,1.719484,1.722022$ & $1/\eta$ & $1.722957$ & C\\
$O$ & $0.2$ & $1600,3200,6400,12800$ & $1.678904,1.682839,1.685877,1.688170$ & $1/\eta$ & $1.688735$ & C\\
$O$ & $0.4$ & $1600,3200,6400,12800$ & $1.576865,1.578771,1.580461,1.581855$ & $1/\eta$ & $1.582037$ & C\\
$O$ & $0.6$ & $1600,3200,6400,12800$ & $1.390192,1.388158,1.387473,1.387107$ & $1/\eta$ & $1.386572$ & C\\
$O$ & $0.8$ & $1600,3200,6400,12800$ & $1.081854,1.070838,1.063533,1.058764$ & $1/\eta$ & $1.056642$ & X\\
$O$ & $0.9$ & $3200,6400,12800,25600$ & $0.833607,0.815763,0.803001,0.793370$ & $1/\eta$ & $0.790598$ & X\\
$P_\pm$ & $1.1$ & $3200,6400,12800,25600$ & $0.838484,0.848199,0.853316,0.854494$ & $1/\sqrt\eta$ & $0.864962$ & X\\
$P_\pm$ & $1.2$ & $1600,3200,6400,12800$ & $0.867311,0.893038,0.909574,0.918839$ & $1/\sqrt\eta$ & $0.948701$ & X\\
$P_\pm$ & $1.3$ & $1600,3200,6400,12800$ & $0.846930,0.872533,0.887675,0.899712$ & $1/\sqrt\eta$ & $0.928644$ & C\\
$P_\pm$ & $1.4$ & $1600,3200,6400,12800$ & $0.790902,0.813704,0.826380,0.840496$ & $1/\sqrt\eta$ & $0.865826$ & C\\
$P_\pm$ & $1.5$ & $1600,3200,6400,12800$ & $0.714961,0.734735,0.746262,0.760383$ & $1/\sqrt\eta$ & $0.782755$ & C\\
$P_\pm$ & $1.6$ & $1600,3200,6400,12800$ & $0.627790,0.642186,0.654805,0.667743$ & $1/\sqrt\eta$ & $0.686890$ & C\\
$P_\pm$ & $1.7$ & $1600,3200,6400,12800$ & $0.529891,0.542042,0.554616,0.565354$ & $1/\sqrt\eta$ & $0.582777$ & C\\
$P_\pm$ & $1.8$ & $3200,6400,12800,25600$ & $0.430790,0.441906,0.450018,0.455005$ & $1/\sqrt\eta$ & $0.468573$ & C\\
\end{tabular}
\end{ruledtabular}
\end{table*}

\begin{table*}[t]
\caption{Component-resolved Dicke extrapolations at $g_x=2$.  Each $N$ entry
denotes four doubling sizes.  The product $K_D^{(\infty)}$ is extrapolated
directly rather than formed from the two component intercepts.  The last
numeric column gives the conservative relative uncertainties of
$(\tlambda_D^{(\infty)},\tOmega_D^{(\infty)},K_D^{(\infty)})$.
The status column uses the C/X convention of Table~\ref{tab:sm-rabi-asymptotic}.}
\label{tab:sm-dicke-asymptotic}
\squeezetable
\scriptsize
\begin{ruledtabular}
\begin{tabular}{lccccccc}
Saddle & $g_p$ & $N$ & $\tlambda_D^{(\infty)}$ & $\tOmega_D^{(\infty)}$
& $K_D^{(\infty)}$ & relative uncertainty (\%) & status\\
\hline
$O$ & $0$ & $160$--$1280$ & $0.999937$ & $1.732125$ & $1.732020$ & $(0.44,0.11,0.44)$ & C\\
$O$ & $0.2$ & $160$--$1280$ & $0.894393$ & $1.897454$ & $1.697072$ & $(0.31,0.16,0.31)$ & C\\
$O$ & $0.4$ & $80$--$640$ & $0.774544$ & $2.049815$ & $1.587671$ & $(0.36,0.32,0.36)$ & C\\
$O$ & $0.6$ & $80$--$640$ & $0.632469$ & $2.191515$ & $1.386064$ & $(0.12,0.33,0.33)$ & X\\
$O$ & $0.8$ & $80$--$640$ & $0.447321$ & $2.325121$ & $1.040083$ & $(0.24,0.18,0.24)$ & X\\
$O$ & $0.9$ & $80$--$640$ & $0.316511$ & $2.390525$ & $0.756707$ & $(0.67,0.18,0.67)$ & X\\
$P_\pm$ & $1.1$ & $80$--$640$ & $0.413316$ & $2.504250$ & $1.035059$ & $(0.41,0.16,0.41)$ & X\\
$P_\pm$ & $1.2$ & $80$--$640$ & $0.534102$ & $2.586915$ & $1.381669$ & $(0.23,0.25,0.25)$ & X\\
$P_\pm$ & $1.3$ & $40$--$320$ & $0.590134$ & $2.698232$ & $1.592241$ & $(0.92,0.64,0.92)$ & X\\
$P_\pm$ & $1.4$ & $40$--$320$ & $0.605207$ & $2.839623$ & $1.718447$ & $(1.00,0.84,1.00)$ & X\\
$P_\pm$ & $1.5$ & $40$--$320$ & $0.589536$ & $3.012375$ & $1.775766$ & $(0.88,1.05,1.05)$ & X\\
$P_\pm$ & $1.6$ & $40$--$320$ & $0.548747$ & $3.217358$ & $1.765370$ & $(0.63,1.25,1.25)$ & X\\
$P_\pm$ & $1.7$ & $40$--$320$ & $0.485751$ & $3.454900$ & $1.678074$ & $(0.28,1.40,1.40)$ & X\\
$P_\pm$ & $1.8$ & $40$--$320$ & $0.399977$ & $3.724676$ & $1.489634$ & $(0.17,1.53,1.53)$ & C\\
$P_\pm$ & $1.9$ & $40$--$320$ & $0.281812$ & $4.025626$ & $1.134357$ & $(1.33,1.68,1.68)$ & C\\
\end{tabular}
\end{ruledtabular}
\end{table*}

The asymptotic scales determine the amplitudes through
Eqs.~\eqref{eq:sm-rabi-invariant} and \eqref{eq:sm-dicke-invariant}, using
the multiplicity identified from the stationary-point topology.
Table~\ref{tab:sm-quadratic-closure} compares the reconstructed map and
amplitudes with independent benchmarks.  In the map column, each entry is
$K_D^{(\infty)}/\tlambda_R^{(\infty)}$ followed by the analytic target in
Eq.~\eqref{eq:sm-cross-model-map}.  In the amplitude
columns, each entry is the dynamical reconstruction followed by the
independent phase-space DOS integration.

\begin{table*}[t]
\caption{Complete quadratic-saddle closure at $g_x=2$.  All scales and
coefficients are dimensionless as written.}
\label{tab:sm-quadratic-closure}
\squeezetable
\scriptsize
\begin{ruledtabular}
\begin{tabular}{cccccc}
$g_p$ & $\tlambda_R^{\rm dyn}$ & $K_D^{\rm dyn}$ & map: dyn./target
& $\omega A_R$: dyn./int. & $\omega\bar A_D$: dyn./int.\\
\hline
$0$ & $1.722957$ & $1.732020$ & $1.00526\,/\,1.000$ & $0.184746\,/\,0.183776$ & $0.183780\,/\,0.183776$\\
$0.2$ & $1.688735$ & $1.697072$ & $1.00494\,/\,1.000$ & $0.188490\,/\,0.187566$ & $0.187564\,/\,0.187566$\\
$0.4$ & $1.582037$ & $1.587671$ & $1.00356\,/\,1.000$ & $0.201203\,/\,0.200516$ & $0.200489\,/\,0.200516$\\
$0.6$ & $1.386572$ & $1.386064$ & $0.99963\,/\,1.000$ & $0.229566\,/\,0.229720$ & $0.229650\,/\,0.229720$\\
$0.8$ & $1.056642$ & $1.040083$ & $0.98433\,/\,1.000$ & $0.301247\,/\,0.306294$ & $0.306043\,/\,0.306291$\\
$0.9$ & $0.790598$ & $0.756707$ & $0.95713\,/\,1.000$ & $0.402619\,/\,0.421612$ & $0.420652\,/\,0.421567$\\
$1.1$ & $0.864962$ & $1.035059$ & $1.19665\,/\,1.210$ & $0.736009\,/\,0.744647$ & $0.615056\,/\,0.615403$\\
$1.2$ & $0.948701$ & $1.381669$ & $1.45638\,/\,1.440$ & $0.671043\,/\,0.663564$ & $0.460761\,/\,0.460804$\\
$1.3$ & $0.928644$ & $1.592241$ & $1.71459\,/\,1.690$ & $0.685537\,/\,0.675466$ & $0.399826\,/\,0.399682$\\
$1.4$ & $0.865826$ & $1.718447$ & $1.98475\,/\,1.960$ & $0.735274\,/\,0.725551$ & $0.370462\,/\,0.370178$\\
$1.5$ & $0.782755$ & $1.775766$ & $2.26861\,/\,2.250$ & $0.813307\,/\,0.805820$ & $0.358504\,/\,0.358142$\\
$1.6$ & $0.686890$ & $1.765370$ & $2.57009\,/\,2.560$ & $0.926815\,/\,0.922086$ & $0.360615\,/\,0.360190$\\
$1.7$ & $0.582777$ & $1.678074$ & $2.87945\,/\,2.890$ & $1.092391\,/\,1.094863$ & $0.379375\,/\,0.378845$\\
$1.8$ & $0.468573$ & $1.489634$ & $3.17909\,/\,3.240$ & $1.358636\,/\,1.381923$ & $0.427367\,/\,0.426519$\\
$1.9$ & --- & $1.134357$ & --- & --- & $0.561216\,/\,0.558378$\\
\end{tabular}
\end{ruledtabular}
\end{table*}

Across the 14 common ARM--ADM couplings, the largest relative deviation from
the map is $4.29\%$.  Across all 29 amplitudes, the largest relative difference
between the dynamical reconstruction and phase-space integration is $4.50\%$.
Restricting the map to couplings classified C in both
models reduces the maximum deviation to $1.88\%$; across the individually C
amplitude sequences, the maximum is $1.69\%$.  No DOS slope, offset, or
model-conversion factor is fitted.  For the Rabi sequences at $g_p=0.9$ and
$1.1$, the largest
$\eta=25600$ response-window spreads are $2.79\%$ and $2.10\%$, and the
paired-cutoff drifts are $0.10\%$ and $0.07\%$.
For the Dicke sequences at $g_p=0.9,1.1,1.9$, the largest window spread is
$1.68\%$ and the largest cutoff drift
is $1.45\times10^{-6}\%$.

At the quartic endpoint, the window is selected directly from the recorded
response, $0.02\leq|\Delta\mathcal C|\leq0.75$.  The Rabi trace is fitted to
$L\tau+D\tau^2$, with the presence of $L$ fixed by the measured initial
covariance.  For Dicke dynamics, the fitted stable-mode oscillation is
resolved separately from the soft component before $D_D$ is extracted.  The
recorded response alone sets this window; $B_R$ and the analytic rate enter
only after extraction.  At $N=320$ and
$\eta=6400$, the amplitude ratios in Figs.~5(c) and 5(d) of the main text are
$1.0004907$ and $1.0005765$, corresponding to deviations of $0.049\%$ and
$0.058\%$ from the analytic endpoint coefficients.

\section{Classical energy-shell chaos protocol}
\label{sec:sm-classical}

The nonsingular Bloch-sphere variables are
$\bm y=(s_x,s_y,s_z,X_c,P_c)$ with $|\bm s|=1$ and
\begin{equation}
h(\bm y)=\frac{s_z}{4}+\frac{X_c^2+P_c^2}{2}
+\frac{g_x}{2}s_xX_c-\frac{g_p}{2}s_yP_c.
\label{eq:sm-classical-h}
\end{equation}

We first derive the origin-saddle barrier used in Fig.~2(b) of the main
text.  On the isotropic line $g_p=0$, minimizing
Eq.~\eqref{eq:sm-classical-h} over the spin orientation gives the lower
energy surface
\begin{equation}
h_-(X_c,P_c)=\frac{X_c^2+P_c^2}{2}
-\frac14\sqrt{1+4g_x^2X_c^2}.
\label{eq:sm-isotropic-surface}
\end{equation}
For $g_x>1$, the origin is a saddle with energy $h_{\rm s}=-1/4$.
The two superradiant minima lie at
$X_c=X_\pm=\pm\sqrt{g_x^4-1}/(2g_x)$ and $P_c=0$, with common energy
\begin{equation}
h_0=h_-(X_\pm,0)
=-\frac18\left(g_x^2+g_x^{-2}\right).
\label{eq:sm-isotropic-ground}
\end{equation}
The barrier between the wells is therefore
\begin{equation}
\begin{aligned}
\Delta h_{\rm sep}
&=h_{\rm s}-h_0\\
&=-\frac14+\frac18\left(g_x^2+g_x^{-2}\right)
=\frac{(g_x-g_x^{-1})^2}{8}.
\end{aligned}
\label{eq:sm-saddle-barrier}
\end{equation}
Writing a fixed-excitation shell as $h=h_0+\delta h$, the shell reaches the
origin saddle only when $\delta h\geq\Delta h_{\rm sep}$.
The equality gives the intersection of the dashed barrier and each fixed
excitation in Fig.~2(b),
\begin{equation}
g_x^*=\sqrt{1+2\delta h}+\sqrt{2\delta h}.
\label{eq:sm-saddle-intersection}
\end{equation}
This condition marks the loss of interwell saddle access.
Finite $f_{\rm ch}$ beyond $g_x^*$ therefore identifies intrawell chaotic
motion after interwell saddle access is lost.

We sample the oriented Poincar\'e section
$s_y=0$, $\dot s_y>0$ uniformly in its canonical area using scrambled Sobol
points.  The Hamiltonian flow and its analytic tangent flow are integrated
together.  Tangent vectors are projected onto both the Bloch-sphere tangent
space and the fixed-energy shell before renormalization.

At fixed $(X_c,P_c)$ the energy constraint on this section has at most two
spin solutions; both are lifted and the orientation condition retains the
positive crossing.  The sampling square extends beyond the numerically
accessible section on every reported shell.  Generic recurrent trajectories
cross this section; tangencies and invariant sets confined to $s_y=0$ have
zero section area.  The reported $f_{\rm ch}$ is therefore the canonical
section-area fraction of the energy-shell return map.  It is not treated as
an independent-binomial sample of the shell volume.

For $N_{\rm ic}$ sampled initial conditions with late-window exponents
$\lambda_i^{\rm FTLE}$, the operational chaotic fraction is
\begin{equation}
\begin{aligned}
f_{\rm ch}&=\frac{1}{N_{\rm ic}}\sum_{i=1}^{N_{\rm ic}}
\Theta\!\left(\lambda_i^{\rm FTLE}-\lambda_c\right),\\
\lambda_c&=0.01.
\end{aligned}
\label{eq:sm-chaotic-fraction}
\end{equation}
Here $\Theta$ is the Heaviside function.  The conclusions are checked
against integration time, time step, sample count, and threshold.  Thus $f_{\rm ch}=0$ means that no sampled exponent
crosses $\lambda_c$ over these resolution tests.

For the same-shell point $(g_x,g_p)=(2,1.4)$,
$h=h_P=-0.3087755102$, we use $T=1600$ and time step $0.01$.
Eight independently scrambled Sobol
replicates, each containing 512 trajectories, give no exponent above the
$0.01$ threshold.  Their median final-window exponents range from
$1.121\times10^{-3}$ to $1.163\times10^{-3}$, while the local Dicke saddle
rate remains $\tlambda_D^{(P)}=0.605556$.

\begin{figure}[t]
\includegraphics[width=\linewidth]{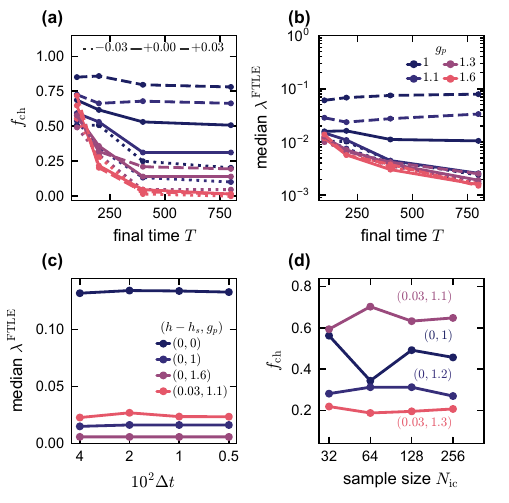}
\caption{Classical convergence at $g_x=2$ and $\lambda_c=0.01$.
(a,b) Chaotic fraction and median FTLE versus final integration time;
colors encode $g_p=1,1.1,1.3,1.6$, and dotted, solid, and dashed lines
encode $h-h_s=-0.03,0,0.03$, with $h_s=-1/4$ the origin energy throughout.
(c) Time-step and (d) section-sample
convergence; labels give $(h-h_s,g_p)$.}
\label{fig:sm-classical-convergence}
\end{figure}

\section{Audit of the fixed-phase paths}
\label{sec:sm-path-audit}

Table~\ref{tab:sm-path-audit} records the endpoints used in
Fig.~2 of the main text.  Every classical point uses $T=800$, time step
$0.01$, transient $50$, threshold $0.01$, and 512 section points.  The two
Path-N spectral endpoints use $N=120$, $n_{\max}=480$, both parity sectors,
and the window $h=0.8\pm0.02$.  The distributions contain 1163 and 1158
adjacent-gap ratios.  Along Path S the thermodynamic photon density is
$0.9375$; at $N=160$ its range over the path is $8.46\times10^{-4}$ and its
largest error from this value is $7.13\times10^{-4}$.

\begin{table*}[t]
\caption{Numerical audit of the two fixed-phase counterexample paths.
$D_P$ and $D_G$ are the Kolmogorov--Smirnov distances to the Poisson and
GOE ratio distributions.}
\label{tab:sm-path-audit}
\begin{ruledtabular}
\begin{tabular}{lccccc}
Case $(g_x,g_p;h)$ & $f_{\rm ch}$ & median FTLE & $\langle r\rangle$ & $D_P$ & $D_G$\\
\hline
N regular $(0.4,0;0.8)$ & $0.00195$ & $0.00176$ & $0.40028$ & $0.0389$ & $0.2550$\\
N chaotic $(0.9,0;0.8)$ & $1.00000$ & $0.11917$ & $0.52578$ & $0.2438$ & $0.0293$\\
S chaotic $(2,0;-0.22)$ & $0.99219$ & $0.20241$ & --- & --- & ---\\
S integrable axis $(2,2;-0.22)$ & $0$ & $0.00173$ & --- & --- & ---\\
\end{tabular}
\end{ruledtabular}
\end{table*}

\section{Parity-resolved spectral statistics on the same shell}
\label{sec:sm-spectrum}

For ordered levels within each parity sector, we form
$r_n=\min(s_n,s_{n+1})/\max(s_n,s_{n+1})$, where
$s_n=E_{n+1}-E_n$ \cite{Atas2013}; ratios are never formed across sectors.
The numerical parity blocks use $J_z+j$, shifting every energy by
$N\Delta/2$ relative to Eq.~(1) of the main text.  We therefore target
$E_{\rm sh}=2N\Delta(h+1/4)$; this constant shift leaves the level spacings
unchanged and selects the physical shell $h=h_P$.

The result in Fig.~4(b) uses $N=400$, $n_{\max}=1600$, and the
window $h_P\pm0.0125$.  We diagonalize the real symmetric Hamiltonian in four
overlapping shift--invert slices within each parity sector.  Each sector
contributes 1888 ratios.  Pooling them only after the within-sector ratios
are formed gives $n_r=3776$ and
$\langle r\rangle=0.393989\pm0.004633$.  The Kolmogorov--Smirnov distances
are $0.0156$ to Poisson and $0.2430$ to GOE, supporting Poisson statistics.
At $N=80$,
the sliced and complete interval spectra contain the same levels and differ
by at most $1.22\times10^{-11}$.

Cutoff tests at $N=120$ use $n_{\max}=360,420,480$ in both parity
sectors.  All six calculations cover the requested window, with converged
sparse eigensolvers; the cutoff/parity spread in $\langle r\rangle$ is
$0.00749$.  Across $N=120,160,200,240$, the mean ratios range from
$0.382$ to $0.397$, consistent with the $N=400$ result.  The larger
$N=400$ sample reduces the statistical uncertainty and clearly favors
Poisson over GOE statistics.

The finite-$N$ Dicke $P$-saddle packet has mean energy $h_P$, an energy
half-width inside the statistics window, and equal total weight in the two
parity sectors.  The displaced ARM packet is centered at the corresponding
common energy $\epsilon_P=2h_P=-0.6175510204$.  Hence the ARM trace, the ADM
packet and classical shell, and the spectral window refer to the same
reduced energy within their stated finite-size resolutions.
The ADM traces in Fig.~4(c) use $N=320$ and 601 points over
$0\leq\tau\leq3$.  Upward cutoff comparisons use $n_{\max}=240,320$ for the
origin packet and displaced-basis cutoffs $n_{\max}=480,640$ for the
$P_\pm$ packet.  The largest displayed-variance cutoff drift is
$1.35\times10^{-12}$ and the largest norm error is $2.34\times10^{-13}$.

\section{Crossover variables and fit-window policy}
\label{sec:sm-crossover}

Near the quartic endpoint $g_p=1$, the Gaussian limit is nonuniform.  The
dimensionless variables
\begin{equation}
\begin{aligned}
u_R&=\frac{\eta(1-g_p^2)^2}
{g_p^2(g_x^2-g_p^2)}, &
u_D&=\frac{N(1-g_p)^2}{g_x-1},
\end{aligned}
\label{eq:sm-crossover}
\end{equation}
compare the closing quadratic-saddle scale with the quantum packet width.
At $g_p=0$, $u_R$ is understood as its limiting value $+\infty$.
In the
origin-side calibration, resolved Rabi fits with $u_R\geq200$ have a maximum
rate error $2.35\%$; Dicke fits with $u_D\geq50$ have a maximum error
$2.62\%$.  These bounds define conservative distances from the endpoint.
The displaced-saddle sequences are also required to pass the response-window,
cutoff, norm, and extrapolation checks described above.
A two-parameter linear extrapolation is classified as asymptotically
controlled when at least three sizes satisfy the corresponding bound, leaving
one residual degree of freedom.  A sequence with fewer than three such sizes
is retained as a crossover-limited large-size extrapolation.  The C/X entries
in Tables~\ref{tab:sm-rabi-asymptotic} and
\ref{tab:sm-dicke-asymptotic} apply this criterion separately to the ARM and
ADM sequences.  The variables do not
define an estimator-independent universal rate curve through the endpoint,
where the correct observable changes from an exponential rate to the
quadratic coefficient in Eq.~\eqref{eq:sm-general-endpoint}.